\documentclass[aps,prl,twocolumn,showpacs]{revtex4}
\newcommand{\bsim}{\mbox{\raisebox{-0.1cm}{$\;
\stackrel{\textstyle>}{\sim}\;$}}}
\newcommand{\lsim}{\mbox{\raisebox{-0.1cm}{$\;
\stackrel{\textstyle<}{\sim}\;$}}}
\usepackage{psfig}

\begin{document}

\title{Relevance of multiband Jahn-Teller effects on the electron-phonon 
interaction in $A_3$C$_{60}$}

\author{E. Cappelluti$^{1,2}$, P. Paci$^2$,
C. Grimaldi$^3$ and L. Pietronero$^2$}

\affiliation{$^1$``Enrico Fermi'' Research Center, c/o Compendio del Viminale,
v. Panisperna 89/a, 00184 Roma, Italy}

\affiliation{$^2$Dipart. di Fisica, Universit\`a ``La Sapienza",
P.le A.  Moro, 2, 00185 Roma,
and INFM RM1, Italy}

\affiliation{$^3$
\'Ecole Polytechnique F\'ed\'erale de Lausanne,
Laboratoire de Production Microtechnique, CH-1015, Lausanne, Switzerland}

\date{\today}

\begin{abstract}
Assessing the effective relevance of
multiband effects in the fullerides
is of fundamental importance to understand the complex
superconducting and transport properties of these compounds.
In this paper we investigate in particular
the role of the multiband effects on the electron-phonon (el-ph) properties
of the $t_{1u}$
bands coupled with the Jahn-Teller
intra-molecular $H_g$ vibrational modes in the C$_{60}$ compounds.
We show that, assuming perfect
degeneracy of the electronic bands, vertex diagrams arising from the
breakdown of the adiabatic hypothesis, are one order of magnitude smaller
than the non-crossing terms usually retained in the Migdal-Eliashberg (ME)
theory.
These results permit to understand
the robustness on ME theory
found by numerical calculations. The effects
of the non degeneracy of the $t_{1u}$ in realistic systems are also analyzed.
Using a tight-binding
model 
we show that the el-ph interaction is mainly dominated by
interband scattering within a single electronic band. Our results
question
the reliability of a degenerate band modeling and show the importance
of these combined effects in the $A_3$C$_{60}$ family.
\end{abstract}
\maketitle

Although there is a large consensus about the phonon-mediated
nature of the superconducting coupling in fullerides $A_3$C$_{60}$
compounds, the precise identification of the origin of so high
critical temperatures ($T_c$ up to $\simeq 40$ K)
is still object of debate.
Electron and phonon properties
in these compounds
are described in terms of molecular crystals,
where the small hopping integral $t$ between
nearest neighbor C$_{60}$ molecular orbitals sets the scale
of the electronic kinetic energy, while 
electron-phonon coupling is essentially dominated
by the intra-molecular vibrational modes \cite{gunnarsson}.
First-principle calculations
suggest that el-ph coupling is mainly dominated
by the $H_g$ modes, having energies $\omega_{H_g} \sim 30-200$ meV,
with some small contribution from the $A_g$ modes.
A simple application of
the conventional Migdal-Eliashberg theory
in these materials is however questioned
for different reasons. On one hand, the small value
of $t$ gives rise to strong electronic correlation
effects since $t$ appears to be of the same order
of the intra-molecular Hubbard repulsion $U$ \cite{gunnarsson}.
On the other hand, nonadiabatic effects are also expected
to be relevant since the electronic energy scale $t$
is of the same order as well of the phonon frequency scale
$\omega_{\rm ph}$ \cite{cgps}.
In addition, the multiband nature of the electronic states $t_{1u}$
involved in the pairing and the Jahn-Teller nature of the
$H_g$ phonon modes make
the problem even more complex but also more interesting.

A popular tool to investigate el-ph properties
in systems where all the energy scales
$t$, $U$, $\omega_{\rm ph}$ are of similar order of magnitude,
is the Dynamical Mean Field Theory (DMFT) \cite{georges}
which does not rely on any small parameter expansion.
In DMFT the lattice Hamiltonian is mapped in a local
impurity model where the electron dynamics through the crystal
is taken into account by a set (in multiband systems)
of frequency-dependent Weiss fields.
This approach is particularly suitable
in fullerides because the intra-molecular el-ph interaction,
just as the Hubbard repulsion, is {\em local} by construction when
expressed in the basis of molecular orbitals. In order to provide
a close set of equations, the hopping integrals $t^{mm'}_{ij}$
are often
chosen to be diagonal with respect to the indexes of
molecular orbitals $m$, $m'$, and, in the absence of crystal field splitting,
degenerate $t^{mm'}_{ij}=\delta_{mm'} t_{ij}$.
By using this approximation, the interplay between electronic
correlation and el-ph coupling mediated by Jahn-Teller mode
has been studied, and a reduction of the effective
superconducting pairing for small $U$ \cite{han} accompanied
by a remarkable enhancement
of it close to the metal-insulator \cite{capone}
transition has been found.
Within the same approximation
the validity of ME theory for a multiband Jahn-Teller 
($t \times H$) model
has also been investigated, showing a good
agreement between ME theory and DMFT results up to
intermediate-large values of the el-ph coupling $\lambda \lsim 1$,
in contrast to the case of a single band system interacting
with a non Jahn-Teller mode ($a \times A$) where
ME theory breaks down for $\lambda \bsim 0.5$ \cite{han}.
The numerical solution of DMFT equations makes hard however
to identify the physical origin of these results.

Aim of this paper is to provide an analytical insight about
the relevance of the multiband and Jahn-Teller-like effects
on the el-ph interaction in fullerides.
In particular we show that, assuming
degenerate electronic bands,
the peculiar symmetry of the Jahn-Teller
$H_g$ modes of the C$_{60}$ molecules gives rise to
a drastic suppression of the el-ph scattering channels
described by vertex diagrams, enforcing the non-crossing
approximation (NCA) which is at the basis of ME theory.
Such a suppression is even more strict in a two band 
$e \times E$ model where vertex diagrams {\em exactly} cancel out.
At the same time, we show that these results strongly depend on the
degenerate band assumption, whereas a careful analysis
of the electronic structure of the $A_3$C$_{60}$
compounds points out that the effective non-degeneracy of the
$t_{1u}$ bands significantly reduces the multiband Jahn-Teller effects.

The simplest way to consider multiband Jahn-Teller effects
in fullerides is to consider
electrons in a tight-binding model on a fcc
lattice interacting with a single five-degenerate $H_g$ phonon mode
with frequency $\omega_0$.
The Hamiltonian reads thus:
\begin{eqnarray}
H^{\rm JT} &=& \sum_{m,m'=1}^3
\sum_{ij} t_{ij}^{m m'}
c_{i m}^\dagger c_{j m'}
+ \omega_0 \sum_{\mu=1}^5 \sum_i
a_{i\mu}^\dagger a_{i\mu}
\nonumber\\
&&
+
\frac{g}{\sqrt{10}} \sum_{\mu=1}^5  \sum_{m,m'=1}^3
\sum_i V_{mm'}^\mu c_{i m}^\dagger c_{im'} ( a_{i\mu}+ a_{i\mu}^\dagger)
,
\label{hamjt}
\end{eqnarray}
where $m,m'$ are indexes of the molecular orbitals, $i$ is the site index
and $\mu$ labels the five-degenerate vibrational modes.
$c$ ($c^\dagger$) and $a$ ($a^\dagger$) are the usual creation (annihilation)
operators for electron and phonon with the quantum number specified
in the Hamiltonian, and the matrices $V_{mm'}^\mu$ are reported
in literature \cite{gunnV}. This model will be compared in the
following with a single band non Jahn-Teller model
defined as:
\begin{equation}
H^{0} = 
\sum_{ij} t_{ij}
c_{i}^\dagger c_{j}
+ \omega_0 \sum_i
a_{i}^\dagger a_{i}
+ g
\sum_i  c_{i}^\dagger c_{i} ( a_{i}+ a_{i}^\dagger).
\label{ham0}
\end{equation}
The prefactor $1/\sqrt{10}$ in the
el-ph scattering term of Eq. (\ref{hamjt}) has been introduced
in order to normalize the effective coupling with
Eq. (\ref{ham0}), namely:
\begin{equation}
\frac{g^2}{10}
\sum_\mu \hat{V}^\mu 
\hat{V}^\mu  = g^2 \hat{I}.
\end{equation}

Let us consider for the moment the case of diagonal and degenerate
hopping integrals $t^{mm'}_{ij}=\delta_{mm'} t_{ij}$, as it
is usually done in DMFT. 
\begin{figure}
\centerline{\psfig{figure=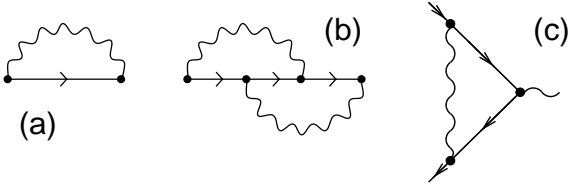,width=8cm,clip=}}
\caption{Skeleton diagrams for: (a)
el-ph self-energy in non-crossing approximation (ME);
(b) lowest order expansion of the el-ph vertex function.
Solid lines are electron Green's functions, wavy lines are phonon propagators,
and small filled circles the el-ph Jahn-Teller matrix elements.
Each pictorial element is a matrix in the space of the
orbital molecular index. The same diagrammatic expressions hold
true even in the superconducting state in terms of Nambu
representation.}
\label{f-self}
\end{figure}
In this case
the molecular orbital index $m$ represents a good quantum number also
for the Bloch-like electron bands, which remain degenerate,
and both electron Green's function $\hat{G}$ and its
self-energy $\hat{\Sigma}$ act thus as the identity matrix $\hat{I}$
in the molecular orbital index $\hat{G} = G \hat{I}$,
$\hat{\Sigma} = \Sigma \hat{I}$.
The el-ph matrix interaction at the ME level (Fig. \ref{f-self}a)
can be evaluated
in the $m \times m'$ ($3 \times 3$) multiband space, and
it reads thus:
\begin{equation}
\hat{\lambda}_{\rm ME}=
\frac{\lambda_0}{10}
\left(
\begin{array}{ccc}
4 & 3 & 3 \\
3 & 4 & 3 \\
3 & 3 & 4
\end{array}
\right),
\label{matrixMEdd}
\end{equation}
where $\lambda_0 = 2 g^2 N(0) / \omega_0 $ is the single
degenerate contribution from intraband and interband scattering
and $N(0)$ is the electron density of states (DOS) for a
{\em single} band of the three-degenerate model).
The effective el-ph coupling in the normal state self-energy
and in the Cooper channels are respectively given by $\lambda_{Z}^m=
\sum_{m'} \lambda_{mm'}=\lambda_0$ and by the maximum
eigenvalue of the matrix (\ref{matrixMEdd}) $\lambda_{\rm SC} =
\mbox{max}_{\rm eig}[\hat{\lambda}]=\lambda_0$ \cite{suhl},
just as in the simple single band analysis.
As a matter of fact, it is easy to see that the matrix
(\ref{matrixMEdd}) can take a block diagonal form and it behaves
as a non Jahn-Teller single band system with effective el-ph coupling
$\lambda_0$ in the one-dimensional symmetrical subspace $\tilde{c} \equiv
1/\sqrt{3}\sum_m c_m^\dagger |0\rangle$.
Note however that some residual el-ph scattering 
with coupling $\lambda_0/10$ is still operative
in the orthogonal non-symmetric space in the molecular
orbital index, although it has no contribution at the
ME level if the band degeneracy is preserved.
As we are going to see, this small component with el-ph coupling
$\lambda_0/10$ is the only
contribution surviving in the higher order terms which
involve vertex diagrams beyond ME theory.

We discuss now the robustness of the ME theory with respect to the inclusion
of highest order diagrams, i.e. vertex corrections,
which are not taken into account in the ME framework. These diagrams are
usually neglected in the ME theory for single band systems
by virtue of the so-called Migdal's theorem which states that
el-ph interaction processes containing vertex
diagrams, as in Fig. \ref{f-self}b, scale with the adiabatic
parameter $\omega_0/t$ and are thus negligible in conventional
materials where $\omega_0/t \ll 1$ \cite{migdal}.
This results does not apply {\em a priori} however in the case
of C$_{60}$ compounds and of other narrow band systems
where $\omega_0 \sim t$ \cite{gpsprl}, so that the agreement between
numerical DMFT data and ME theory remains somehow surprising.
As we are going to show, the physical origin of
such agreement stems from the particular matrix structure of the
$\hat{V}$ matrices in the Jahn-Teller multiband case
of the fullerides
which leads to an almost complete cancellation of
the vertex diagrams independently of the value of the
adiabatic parameter $\omega_0/t$.

Once more, the simplest way to understand this feature
is to assume a diagonal-degenerate hopping term.
In order to clarify the role of multiband Jahn-Teller effects
on the vertex processes, let us compare at the skeleton level, for both
$t \times H$ and $a \times A$ models,
a self-energy contribution involving a vertex diagram 
(Fig. \ref{f-self}b) and a typical vertex-free diagram which is usually
taken into account in ME theory (Fig. \ref{f-self}a).
For a $a \times A$ model we can write in a compact form:
\begin{eqnarray}
\Sigma_{\rm a}^{0}(k) &=& - g^2 \sum_p D(k-p)G(p),
\label{sa0}
\\
\Sigma_{\rm b}^{0}(k) &=& - g^4 \sum_p D(k-p)G(p)\Lambda(p,k),
\label{sb0}
\end{eqnarray}
where $p$ is the momentum-frequency vector (${\bf p},\omega_p$)
and $\sum_p = \sum_{\bf p} \int d\omega_k / 2\pi$, and where
$\Lambda(p,k)$ represents the lowest order vertex correction
depicted in Fig. \ref{f-self}c.
In the early 60's Migdal was able to show that
\begin{equation}
\lim_{\omega_0/t \rightarrow 0}
\Lambda(p,k) \propto  \frac{\omega_0}{t},
\end{equation}
which allows to neglect vertex diagrams in the adiabatic limit
$\omega_0/t \ll $.

The same diagrammatic picture (Fig. \ref{f-self}) and similar
analytical expressions hold true for the multiband Jahn-Teller
$t \times H$ model, properly generalized in the matricial space
of the molecular orbital index. For the
diagonal-degenerate hopping term case the electron Green's function
are once more proportional to the identity matrix $\hat{I}$, so that
the only matricial structure comes from the $\hat{V}$ matrices
in the el-ph scattering term.
In explicit way we can write:
\begin{eqnarray}
\hat{\Sigma}_{\rm a}^{\rm JT}(k) &=& - \frac{g^2}{10} 
\left[\sum_\mu \hat{V}^\mu \hat{V}^\mu \right] \sum_p D(k-p)G(p)
\nonumber\\
&=& 
- g^2 \hat{I}\sum_p D(k-p)G(p),
\label{sajt}
\\
\hat{\Sigma}_{\rm b}^{\rm JT}(k) &=& - \frac{g^4}{100} 
\left[\sum_{\mu \nu} \hat{V}^\mu \hat{V}^\nu 
\hat{V}^\mu \hat{V}^\nu \right]
\nonumber\\
&&\times
\sum_p D(k-p)G(p)\Lambda(p,k)
\nonumber\\
&=&
- \frac{g^4}{10} \hat{I}\sum_p D(k-p)G(p)\Lambda(p,k),
\label{sbjt}
\end{eqnarray}
where the last relation comes from the
matricial properties of the $\hat{V}$ terms:
$[\sum_{\mu \nu} \hat{V}^\mu \hat{V}^\nu 
\hat{V}^\mu \hat{V}^\nu]/100 = \hat{I}/10$.

The comparison between Eqs. (\ref{sa0})-(\ref{sb0})
and Eqs. (\ref{sajt})-(\ref{sbjt}) shows that a strong reduction
of the vertex diagrams,
of a factor 10, is operative in the
$t \times H$ model appropriate for fullerides compounds
in the case of a diagonal-degenerate hopping term,
validating at a large extent the non-crossing approximation
well beyond the adiabatic regime.
We would like to stress that the robustness of the ME theory
with respect to the inclusion of vertex diagrams does not stem
from the negligibility of the vertex function $\Lambda$, which
can be well sizable for $\omega_0 \sim t$, but from the non commutativity
of the $\hat{V}$ matrices and from the assumption of
diagonal and degenerate hopping term. As matter of fact, these
results hold true even if the assumption of a diagonal hopping
term is relaxed as long as the three electronic bands are degenerate.
To show this, we can diagonalize the hopping term in Eq. (\ref{hamjt})
by the transformation:
\begin{equation}
c_a({\bf k}) = \sum_{i,m} M^{am}_{\bf k} c_{i,m}
\mbox{e}^{i {\bf k}\cdot {\bf R}_i},
\label{tras}
\end{equation}
so that the el-ph scattering term becomes:
\begin{equation}
g \hat{V}^\mu
\rightarrow \hat{g}^\mu_{{\bf k},{\bf k+q}}
=
g \hat{U}^\mu_{{\bf k},{\bf k+q}}
=
\hat{M}^{-1}_{\bf k}
\hat{V}^\mu
\hat{M}_{\bf k+q}.
\end{equation}
Since the electron Green's function are still diagonal and degenerate
$\hat{G}(k) = \hat{I} G(k)$ in this new basis,
the matricial structure of the self-energy terms is still only given
by the el-ph matrices $\hat{U}$. 
We have thus
\begin{eqnarray}
&&
\frac{1}{100}
\sum_{\mu \nu} 
\hat{U}^\mu
\hat{U}^\nu
\hat{U}^\mu
\hat{U}^\nu
\nonumber\\
&=& \frac{1}{100}
\sum_{\mu \nu} 
\hat{M}^{-1} \hat{V}^\mu \hat{M}
\hat{M}^{-1} \hat{V}^\nu \hat{M}
\hat{M}^{-1} \hat{V}^\mu \hat{M}
\hat{M}^{-1} \hat{V}^\nu \hat{M}
\nonumber\\
&=&\frac{1}{100}
\hat{M}^{-1} \hat{V}^\mu  \hat{V}^\nu  \hat{V}^\mu  
\hat{V}^\nu \hat{M}
=\frac{1}{10}\hat{I}.
\end{eqnarray}
Finally, it is easy to check that, if the double-degenerate
Jahn-Teller $e \times E$ model is considered, the
non-commutativity of the corresponding $\hat{V}$ matrices leads
to a {\em complete} cancellation of the vertex diagrams.

\begin{figure}
\centerline{\psfig{figure=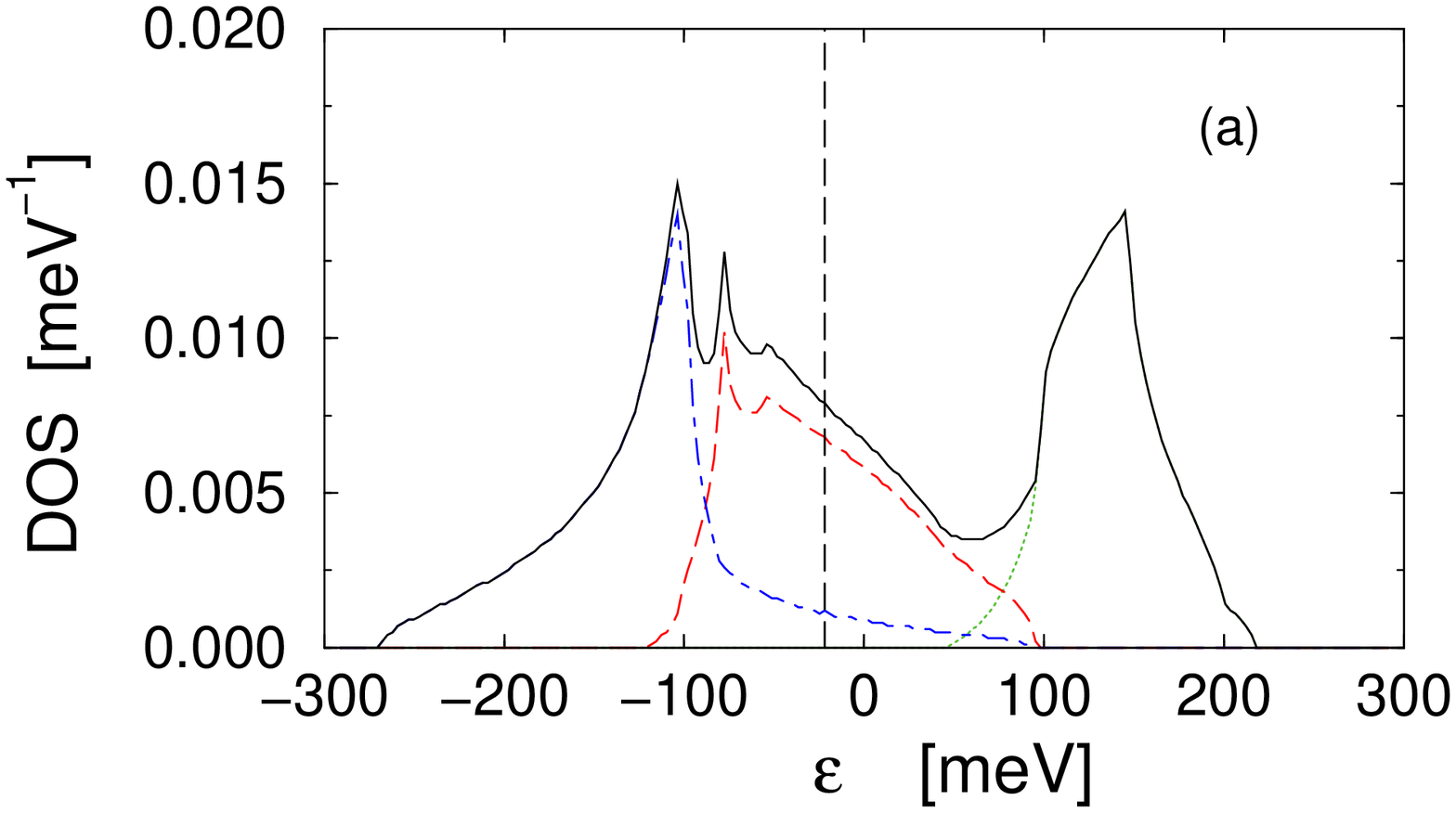,width=8cm,clip=}}
\centerline{\psfig{figure=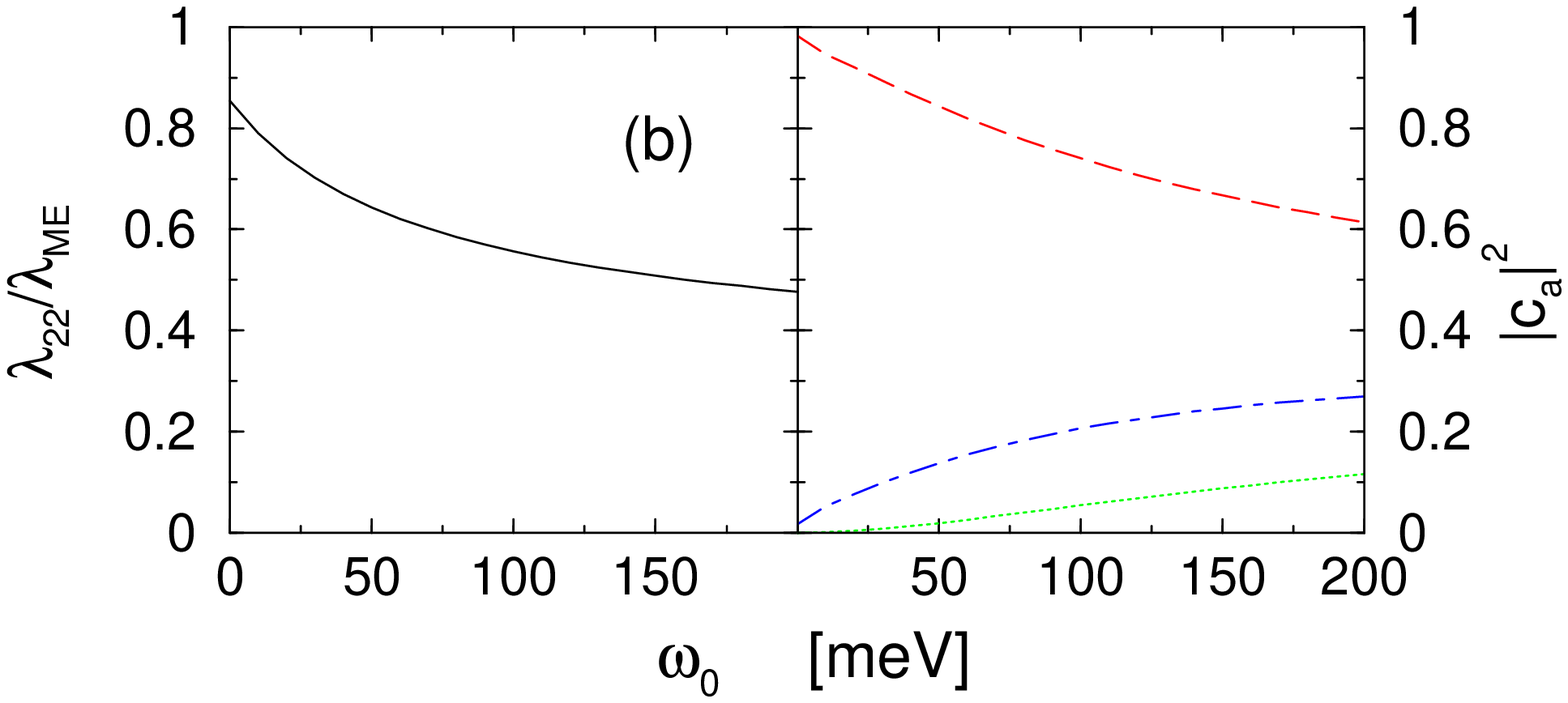,width=8.5cm,clip=}}
\caption{(color online). (a) Total DOS (black solid line) and
partial DOS for each electronic band in the one-directional
TB model for $A_3$C$_{60}$. The
dashed vertical line represents the Fermi level $\mu$. (b) Left panel:
Ratio between
the only intraband contribution $\lambda_{22}$ and the total el-ph
coupling $\lambda_{\rm ME}$. Right panel: band components $|c_i|^2$
of the eigenvector of the multiband el-ph coupling matrix $\hat{\lambda}$.}
\label{f-dos}
\end{figure}

From the above discussion, one could be tempted to conclude that
an effective non-crossing ME theory is expected to be enforced
in fullerides due to an almost complete cancellation of
the vertex processes. However, it should be stressed the above results
are valid as long as the electronic bands can be considered degenerate.
In the last part of this paper, we are going to argue that this
latter assumption is not appropriate in $A_3$C$_{60}$ compounds
in regards with el-ph properties,
and a single band model, which is unaffected by Jahn-Teller
effects, is a better starting point for $\omega_0 \ll t$.
In order to introduce a more realistic electronic band structure
than a simple diagonal-degenerate model in the molecular orbital space,
we consider a tight-binding (TB) model for
the $t_{1u}$ bands as discussed
in Ref. \cite{satpathy} which reproduces first-principle LDA calculations
with a high degree of precision.
For sake of simplicty we consider the one-directional TB model
where all the C$_{60}$ molecules have a fixed orientation in the fcc cystal.
Different directional ordering will be discussed
in a more extended publication but they are not expected to 
qualitatively affect
our results since they still predict a set of narrow bands
with only few Fermi cuts.
The electronic hopping term in Eq. (\ref{hamjt}) is then diagonalized
by the $\hat{M}$ matrix introduced in (\ref{tras}), where
each matrix element depends on the electronic momentum ${\bf k}$.
For a generic point of view,
the degeneracy of the electronic states
in the crystal structure
is preserved only on special high-symmetry points.
The electronic band structure 
along the high-symmetry axes of the fcc Brillouin zone
and the total electron DOS has been
widely reported in literature \cite{gunnarsson}.
In order to assess to which extent
the realistic band structure can be approximated
by a degenerate three-band model,
it should be noted that (low energy) electronic and transport properties
are mainly determined by quantities defined at the Fermi level,
as the Fermi surface itself, the electron DOS at the Fermi level,
the Fermi velocity, etc.
A realistic band structure can thus be modeled as a degenerate
three-band system only if it presents qualitatively similar
Fermi sheets, with similar densities of states and similar Fermi velocities.
In this perspective $A_3$C$_{60}$ compounds
seems not to be a good candidate since it is known to have
only two Fermi cuts of the electronic bands
with quite different Fermi surfaces \cite{FS}.
The smooth topological evolution of the Fermi surfaces with
the Fermi level permits to identify in an unambiguous way,
for what concerns electronic and transport properties,
the three electronic bands with their corresponding DOS.
The electronic DOS for each band, as well as the total DOS,
is plotted in Fig. \ref{f-dos}a, which shows that the electronic
band structure is largely dominated at the Fermi level $\mu = -23$ meV,
by a single band
which is roughly half-filled, with two additional bands which are
quite far from the Fermi level. Only two bands have finite DOS
at the Fermi level, in agreement with the report of only two Fermi surface.
We would like to stress
that the non-degeneracy of the DOS is not related
to the removing of the crystal field symmetry, but it is uniquely 
determined by non degeneracy of the electronic dispersion.
If we evaluated the mean value $\bar{\epsilon}=\langle \epsilon \rangle$
and the variance 
$\delta \epsilon = \langle (\epsilon - \bar{\epsilon})^2 \rangle^{1/2}$
of each band, we obtain
$\bar{\epsilon}_1 = -120$ meV, $\delta \epsilon_1 = 60$ meV,
$\bar{\epsilon}_2 = -23$ meV, $\delta \epsilon_2 = 47$ meV,
$\bar{\epsilon}_3 = 135$ meV, $\delta \epsilon_3 = 30$ meV.
This considerations suggest $A_3$C$_{60}$ can be modeled in
good approximation as a single band system
and that multiband effects are 
qualitatively irrelevant
as far as the phonon energy is not sufficiently high
to switch on el-ph scattering between different bands.

We sustain this intuitive argumentations with a numerical calculation
of the multiband el-ph matrix interaction
$\hat{\lambda}_{\bf k}$
within the non-crossing approximation
(Fig. \ref{f-self}a) which generalizes the ME theory to
systems with generic DOS $N(\epsilon)$.
For sake of simplicity we consider unrenormalized electron and
phonon propagators, with energy spectra respectively given
by the TB electron dispersion and by the phonon frequency $\omega_0$.
A ${\bf k}$-independent el-ph interaction is obtained in the usual
way by a proper average over the Fermi surface:
$\lambda^{ab} = [\sum_{\bf k} D(\epsilon_{\bf k}^a)\lambda_{\bf k}^{ab}]
/[\sum_{\bf k} D(\epsilon_{\bf k}^a)]$, where
$D(\epsilon)=(1/\pi)\omega_0/(\omega_0^2+\epsilon^2)$, which
reduces to $\sum_{\bf k} D(\epsilon_{\bf k}^a) \simeq N(\epsilon_{\rm F}^a)$
for $\omega_0 \rightarrow 0$.
For sake of simplicity we neglect self-energy effects, which are
expected to shrink furthermore the electron bands at the Fermi level.
In order to provide a 
first simple estimate of multiband effects,
in the left panel of Fig. \ref{f-dos}b we compare
the effective total el-ph coupling $\lambda_{\rm ME}$,
obtained as the maximum
eigenvalue of the matrix $\hat{\lambda}$, with the single intraband
contribution $\lambda_{22}$ relative to the high DOS
central band shown in Fig. \ref{f-dos}a.
For $\omega_0 \rightarrow 0$, $\lambda_{22}/\lambda_{\rm ME} \simeq 0.86$,
pointing out that the dominant el-ph scattering comes from intraband
processes with a small interband contribution due to
the small DOS of the lowest electron band. As expected the discrepancy 
between  $\lambda_{22}$ and $\lambda_{\rm ME}$ becomes
more relevant as the activation of high energy interband processes
by the phonon increases. To quantify further the role of multiband
effects we plot in the right panel of Fig. \ref{f-dos}b
the components of the eigenstate of the el-ph scattering matrix
on the Bloch-like band index $a$. For 
$\omega_0 \rightarrow 0$ only the $|c_2|^2$ component is sizable
reflecting that el-ph scattering is mainly intraband. By increasing
$\omega_0$ interband processes are gradually turned on, but, for
the physical range of $\omega_0 \lsim 200$ meV, intraband el-ph
scattering within the central band are still dominant. In the opposite limit
$\omega \gg W$ ($W$ being tht total electronic bandwidth) 
the three components become
identical and the $A_3$C$_{60}$ compounds are expected to
behave has a degenerate three band system.

In conclusion, in this paper
we have investigated the role of the electronic structure on the multiband
Jahn-Teller effects in the el-ph interaction in fullerides compounds.
We found that a band degenerate model leads to an almost
cancellation of vertex diagrams enforcing the validity of ME theory,
as observed by Quantum Monte Carlo DMFT data.
On the other hand, we also show that realistic band structure
calculations suggests that the el-ph interaction
in $A_3$C$_{60}$ compounds is mainly dominated by intraband
scattering within a single band. Note that similar conclusions are not
expected to
apply for electron-electron (Hubbard) interaction where relevant
scattering processes are not restricted in an energy window
around the Fermi level.
Note also that electronic and
vibrational disorder could in principle strongly
mix intra- and interband el-ph scattering.

The effective role of the non degeneracy on the
metal-insulator transition driven by the Hubbard
repulsion and the inclusion of 
possible crystal splittings and disorder
will be the future developments
of the present work.

We thank M. Capone and N. Manini for fruitful discussions
and the careful reading of the manuscript.
This work was partially funded by the INFM project PRA-UMBRA
and by the MIUR project FIRB RBAU017S8R.

\end{document}